\documentclass[acmsmall]{acmart}
\usepackage{graphicx}
\usepackage{soul}
\usepackage{enumitem}

\begin{document}

\title[Exploring UI/UX Designers' Privacy Advocacy in Practice]{``We Wanted to Do Better Than the Law'': Exploring UI/UX Designers' Privacy Advocacy in Practice}

\author{Keyu Yao}
\orcid{0009-0003-5709-4175}
\email{keyu.yao@mail.mcgill.ca}
\affiliation{%
 \institution{McGill University}
 \city{Montreal}
 \state{Quebec}
 \country{Canada}}
 
\author{Jinghui Cheng}
\orcid{0000-0002-8474-5290}
\email{jinghui.cheng@polymtl.ca}
\affiliation{%
 \institution{Polytechnique Montreal}
 \city{Montreal}
 \state{Quebec}
 \country{Canada}}

 \author{Jin L.C. Guo}
\orcid{0000-0003-1782-1545}
\email{jguo@cs.mcgill.ca}
\affiliation{%
 \institution{McGill University}
 \city{Montreal}
 \state{Quebec}
 \country{Canada}}

\begin{abstract}
Designers hold primary responsibility for shaping the user interface (UI) and user experience (UX) of a product. This role goes beyond aesthetics and usability, extending to the privacy outcomes of user experience, which often emerge through collaboration with other stakeholders such as developers, product managers, and marketing teams. Previous studies on enhancing privacy for technological products primarily focused on the roles of developers---understanding their needs and challenges---but limited effort is devoted to examining how UI/UX designers consider and approach privacy in their work. Through 12 semi-structured interviews with privacy-advocating UI/UX designers, we explore the perceptions, influencing factors, challenges, and adaptive methods they use regarding privacy implementation. We pay special attention to how these challenges and adaptations play out in team-based settings where decisions are negotiated together. Our study reveals how personal and contextual factors shape designers' value of privacy, the collaborative nature of the challenges designers face when trying to prioritize privacy, and how they navigate tensions between business goals, team dynamics, and technical development. Based on our findings, we discuss implications for advocating a user-centered approach for supporting privacy-aware design, suggestions for organizational-level changes and bridging knowledge gaps through designer-centric tools and community building.

\end{abstract}




\maketitle

\section{Introduction}
As individuals interact with increasingly complex digital applications and services, user privacy issues have become more important than ever~\cite{KOKOLAKIS2017}. The pervasiveness of surveillance capitalism in the technology landscape has exacerbated this problem, with large companies profiting from users' personal data~\cite{zuboff2023age}. To mitigate this issue, regulations have been proposed or enacted in various regions, such as the European Union's General Data Protection Regulation (GDPR)~\cite{gdpr_eu_2016}, California's Consumer Privacy Act (CCPA)~\cite{ccpa}, and Canada's Consumer Privacy Protection Act (CPPA)~\cite{cppa}. These regulations had a dual impact. While they prompted companies to reassess and refine their user data practices, they also inadvertently established a narrow framework for privacy considerations, leading companies to focus on minimal compliance rather than embracing a more comprehensive approach to user privacy. As \citet{waldman2021industry} put, ``With compliance focused on paper trails, checkboxes, and prefilled reports from outsourced vendors, privacy professionals come to confuse mere symbols of compliance with actual adherence to privacy law.''~\cite[p.9]{waldman2021industry}

Ideally, serving as users' primary allies within companies, UI/UX designers should play a critical role in companies and organizations to shape how users' data is handled in technologies. Through design choices, they can influence what information to collect, how transparent the process is, and how users can exercise control over their data. In practice, however, these decisions are made in a collaborative manner in which UI/UX designers have to balance competing demands from other stakeholders, such as business objectives, technical feasibility, and regulatory compliance, when attempting to implement privacy-enhanced options. This creates ``ethical tensions'' for designers, who must navigate between their moral and professional obligations~\cite{ethicalTensions} and the interests and needs of other team members~\cite{Evaluating_Privacy_Perceptions}. Supporting designers' privacy-related design practices in a collaborative environment is thus important for fostering a user-centered approach to data handling and enhancing the overall privacy and trustworthiness of digital products. To provide this support, it is essential to gain a nuanced understanding of how designers' privacy-related values, challenges, and practices intersect with their existing organizational and collaboration contexts.

However, the existing privacy literature tends to focus on technical development rather than UI/UX design and its impact on products' privacy practices. For example, the concept of Privacy by Design mostly emphasizes software architecture and data flow design~\cite{PrivacyDesignStrategies}. Prior work on practitioners' perspectives and approaches to privacy also mostly focused on software developers (e.g., ~\cite{Hadar2018,privacy_champions,Evaluating_Privacy_Perceptions,FactorsAffectingPrivacy}), with very limited attention paid to UI/UX designers~\cite{Discount}. While some studies have explored responsible design practices among UI/UX designers, they have primarily concentrated on manipulative design techniques~\cite{Gray_dark_pattern_18,navigating_gray} or methods for ethical design in general~\cite{Chivukula3}, without thoroughly examining the broader contextual influences in collaborative environments on designers' privacy values, decision-making processes, and practical challenges.

We address this knowledge gap in this paper. Particularly, we aim to understand the mindsets and influencing factors shaping the practices of UI/UX designers concerning privacy, and examine how they navigate challenges in integrating privacy considerations into their innately collaborative design processes. Through 12 semi-structured interviews with privacy-advocating professional designers from diverse industries and regions, our findings revealed important information regarding how those designers form their privacy-related value systems, navigate organizational constraints and tensions, and develop adaptive strategies to incorporate privacy considerations and address those tensions in collaboration. We found that the core values of our participants regarding privacy are both moral and practical, which are influenced by diverse personal and contextual factors. Moreover, the most significant challenges they face are collaborative in nature, from interactions with various stakeholders who have more power and resources over designers and may prioritize business goals and technical development over privacy concerns.

Overall, our study addresses a notable gap in the existing literature related to privacy practices among UI/UX designers in their collaborative processes. We provided empirical evidence and practical insights for future efforts aimed at cultivating privacy advocacy in designers, empowering them in software teams to adopt user-centric data protection approaches, and allowing them to collectively drive technological advances that are privacy-sensitive and reclaim their role in user-focused innovations.

\section{Related Work}

\subsection{Privacy Concerns in Technology}
The dominance of surveillance capitalism has resulted in a substantial portion of the digital economy depending on the collection and the use of personal information from users~\cite{zuboff2023age}. Web platforms, from social networks to e-commerce, rely heavily on user data to optimize functionality and personalize experiences, which can violate the autonomy of users~\cite{social_media_deletion}. 

In addition, unauthorized data sharing, insufficient data protection, and opaque data-handling practices undermine user trust in contexts where companies employ third-party trackers to monitor browsing behavior without explicit user consent~\cite{crumbled_cookies, user_tracking}. Such practice violates privacy regulations like the European Union's General Data Protection Regulation (GDPR)~\cite{gdpr_eu_2016} and California's Consumer Privacy Act (CCPA)~\cite{ccpa}. At the same time, large companies have prompted increasing privacy and data protection concerns, resulting in regulatory responses such as the above legal legislation~\cite{BusinessRisks}. However, the framings of privacy concerns are mostly as business risks rather than ``a positive good to provide to users''~\cite{BusinessRisks}. For smaller businesses, especially those without an established focus on security, they might lack the necessary resources to cope with the compliance overhead~\cite{are_we_there_yet}. As a result, they might take ```necessary shortcuts' to comply with data protection and outsource some of their duties to third parties''~\cite{Discount}. In complex platform ecosystems, on the other hand, users often face difficulty accurately assessing what data is collected, who it is shared with, and how it is used~\cite{Mental-Model-Based-Facebook,solove}. In this general technological landscape, UI/UX designers are forced to navigate ``\textit{ethical tensions}'' as they balance delivering value to users through engagement with the use of manipulative design strategies~\cite{ethicalTensions}.

Prior research has investigated various manipulative and ethically ambiguous UI/UX design strategies that compromise users' privacy and their autonomy in privacy-related decision-making. Among these, the most extensively studied are known as \emph{dark patterns}~\cite{Gray_dark_pattern_18,DarkPatterns, DarkPatternsGDPR}. Foundational work by \citet{Mathur} categorized dark patterns in e-commerce sites, showing how they exploited cognitive biases to serve business goals at the expense of user autonomy. Follow-up studies refined these definitions through normative considerations~\cite{Mathur2} and concept-based frameworks to evaluate when a design deviates from user expectations to benefit service providers~\cite{dark_patterns_user_expectations}. Research has shown the prevalence of dark patterns across modalities, with many services containing at least one dark pattern~\cite{Modalities}. Similarly, \citet{Asshole} found that a large proportion of ``asshole design'' contains recognized dark pattern strategies, and proposed a theoretical framework to distinguish between different types of problematic design based on designer intent, arguing that designers actively inscribe values into their work.

\subsection{UI/UX Design for Privacy Awareness and Protection} 

Researchers and practitioners have explored various design efforts aimed at guiding users toward safer, more informed choices regarding their privacy awareness and protection. In direct response to the proliferation of dark patterns, \citet{dark_patterns_user_expectations} proposed an alternative view focused on standard software concepts and used this framework to encourage design that facilitates positive expected behavior. Other behavioral and structural design tactics were also investigated. For example, granular permission controls such as sliders and toggles for categories like location, contacts, or analytics, instead of all‑or‑nothing check boxes, were found to be effective in giving users fine‑grained agency over their data~\cite{Slider}. Strategic friction that inserts small obstacles or delays before users can complete sensitive actions was also found to promote reflection rather than rote consent~\cite{friction}. Likewise, digital nudges and reflection prompts were investigated to re-frame consent choices to alter privacy attitudes~\cite{Zhang}, encourage deliberation~\cite{nudge}, and change user behavior~\cite{your_location}. Just‑in‑time notices that appear at the exact moment data is collected or shared were also explored to prompt children to pause and reflect before consenting~\cite{Notice_children}. Adopting protection motivation theory, \citet{Shiri_PMT_24} found that UI/UX design can effectively integrate the various considerations of privacy to motivate participants to attend to privacy issues. Moreover, participatory methods such as co‑design workshops and speculative ``privacy fictions'' were also explored to engage end users early to surface expectations and latent threats, building empathy for potential misuse scenarios and informing preventative design~\cite{Merrill_security_fiction_20,EmpowermentLu}.

\subsection{Practitioners' Approaches and Perspectives on Privacy}

Privacy considerations in software teams and the software life cycle have gained increasing attention in recent years~\cite{nistssdf,Spiekermann2012}. For example, through empirical studies with developers and analysis of developers' forums, \citet{Evaluating_Privacy_Perceptions} found that software teams often lack formal privacy training, rely on self-learning, and need role-specific support to handle multi-jurisdictional compliance. Through an ethnographic study on how privacy and security issues are addressed in a software company, \citet{ethnographic} identified that security problems in software often stem from complex stakeholder dynamics rather than developer-only errors and advocated direct collaboration among stakeholders. A narrative review also found that factors like organizational culture, management support, and resource allocation all shaped how privacy is prioritized or sacrificed in software development~\cite{FactorsAffectingPrivacy}. Moreover, a recent review of studies investigating the Privacy by Design approach in software teams revealed a lack of models, processes, and tools to support practitioners in adopting this approach~\cite{PrivacyByDesignSoftwareEngineering}.

Previous work has also explored developers' perceptions and practices towards privacy in their work. For example, \citet{Hadar2018} found that developers relied on security language that narrowed their view of privacy to external threats, were strongly influenced by their organization's privacy climate, and tended to favor policy-based privacy solutions. On the other hand, \citet{privacy_champions} interviewed ``Privacy Champions'' in software teams and found that these individuals often struggled against a negative privacy culture, internal prioritization tensions, and limited tooling, relying on informal discussions and management advocacy to keep privacy on the agenda. The resulting reality is that discourse around privacy often perpetuates organizational power---making the privacy compliance narrowly centred around technical work and the notion of notice and consent, and fails to provide meaningful protection to users~\cite{waldman2021industry}.

Compared to developer-focused studies, far less is known about the practices, perspectives, and challenges UI/UX designers have when considering privacy and embedding privacy into their work. Notable exceptions include a study by \citet{navigating_gray}, which explored designers' perceptions of privacy dark patterns. The authors found factors such as business pressures and design conventions influenced designers' attitudes, often leading them to justify manipulative designs as standard practice. Another relevant study by \citet{Discount} examined the perspectives of users, designers, and business leaders of smart home applications towards data protection. The authors found that smart home designers often faced challenges when aligning privacy considerations with business goals and relied on ``discount data protection''---shortcuts, heuristics, and common-sense solutions---to approach data protection design.

While valuable, these studies are limited in scope, focusing on either a particular design aspect (i.e., dark patterns) or a specific application domain (i.e., smart home). Moreover, they did not explore the broader contextual factors that shape designers' value systems and decision‑making approaches concerning privacy. Our study addresses these gaps by examining how UI/UX designers form privacy-related values, negotiate organizational tensions and practical challenges, and adapt their practices to approach privacy design.

\section{Methods}
To investigate how privacy-advocating designers perceive and implement privacy considerations in their collaborative work environments, we conducted semi-structured interviews with 12 experienced designers who had prior experience attending to privacy in their work. The study was approved by the research ethics committees of all involved institutions.

\subsection{Recruitment and Participants}
\begin{table}[t]
\centering
\caption{Demographics of the Interview Participants}
\label{tab:demographics}
\resizebox{\textwidth}{!}{
\begin{tabular}{lllll}
\toprule
\textbf{ID} & \textbf{Company/Client Type} & \textbf{Current Job Title} & \textbf{Country} & \textbf{Years of Exp.} \\
\midrule
P1 & Healthcare and government consulting & UI Consultant & Australia & 5–10 yrs \\
P2 & Freelance design and coaching & UX Consultant \& Coach & Netherlands & $>$20 yrs \\
P3 & E-commerce & UI/UX Designer & Germany & 5–10 yrs \\
P4 & Digital agency & Intermediate Experience Designer & Canada & $<$5 yrs \\
P5 & Software/Internet & Senior UX Designer in Privacy \& Security & USA & 10-20 yrs \\
P6 & B2B product & UI/UX Designer & Canada & $<$5 yrs \\
P7 & Marketing \& digital agency & UI/UX Designer Intern & USA & $<$5 yrs \\
P8 & Freelance/Government & UI/UX Designer & Netherlands & $<$5 yrs \\
P9 & Freelance & Graphic Designer & Kazakhstan & 5-10 yrs \\
P10 & Freelance & Product Manager \& UX Designer & India & 10-20 yrs \\
P11 & Product company & UI/UX Designer & India & $<$5 yrs \\
P12 & AI-powered content service & Service/Industrial Designer & India & $<$5 yrs \\
\bottomrule

\end{tabular}
}
\end{table}

We aimed to recruit professional UI/UX designers who have had experience attending to privacy during their design process. Our recruitment took place in two rounds to achieve a diverse sample across industries, levels of experience, and geographic regions.
In the first round, we conducted a web search using the keywords ``Privacy'' and ``UI/UX design'' on the DuckDuckGo search engine. From this, we identified individuals who had published privacy-related articles or blogs online (e.g., Medium, UX Collective). We then filtered these individuals to identify those who had practical experience as UI/UX designers, either currently or in the past, based on their LinkedIn profiles. We then reached out to eligible individuals via email or LinkedIn. We were able to recruit five blog authors as participants during this round.
In the second round, we expanded recruitment through a combination of advertisements and personal networks. A public call for participation was shared on LinkedIn, targeting design professionals who are interested in privacy-related topics, and was further circulated through the researchers' personal contacts in the industry. Through this round, we recruited seven additional participants. 

This strategy enabled us to recruit 12 designers in total, working across freelance, in-house, and agency settings, with representation from different regions, including North America, Europe, Asia, and Australia. Participants varied in experience (ranging from two to over 20 years) and in their exposure to privacy-related design challenges. Table~\ref{tab:demographics} provides an overview of our participants. 

\subsection{Interview Process}
We developed the interview questions to prompt reflection on the participant's design practices, organizational culture, and collaborative and personal relationship to privacy. The interview guide includes four main parts as follows:

\begin{enumerate}[topsep=0pt]
  \item \textit{Roles \& Responsibilities}: Participants were asked to walk through a typical design workflow, outline their core responsibilities, and describe how they collaborate with peers and stakeholders to achieve a specific task.  
  \item \textit{Privacy Perceptions}: Participants were asked to articulate their own definition of privacy and explain where, why, how, and for whom it matters in their practice.  
  \item \textit{Privacy‑Oriented Design Practices}: Participants were asked to reflect on past projects and detail the concrete methods, tools, or heuristics they used to integrate privacy practices into UI/UX design.  
  \item \textit{Challenges \& Adaptive Strategies}: Participants were asked to identify obstacles they face when incorporating privacy and outline the negotiation or workaround strategies they employed.  
\end{enumerate} 
The interviews were recorded and then fully transcribed. All identifying information about the participants was anonymized in the final transcripts. Each participant was rewarded \$30 CAD for their time.

\subsection{Data Analysis}
The transcripts of the interviews were analyzed iteratively using the reflexive thematic analysis approach~\cite{Braun2021}, allowing themes to emerge directly from the data rather than being pre-imposed. The analysis proceeded through multiple stages and was guided by our focus on understanding designers' privacy-related values, practices, and challenges in collaborative settings. First, researchers independently conducted open coding by reading through transcripts and identifying preliminary codes based on recurring concepts. These codes were then discussed collaboratively and refined into broader conceptual categories. Next, the researchers further grouped the codes, re-reading the transcripts when necessary, to identify core themes and ensure consistency across the dataset. The iterative process ensured both depth and reliability in capturing the nuanced perspectives of UI/UX designers on privacy.
\section{Results} 
In this section, we describe our participants' core values regarding privacy, the social and collaborative factors that shaped these values, the challenges they face when they navigate organizational tensions to incorporate privacy, and the adaptive strategies they employ to address these challenges.

\subsection{Core Values Designers Held Concerning Privacy}
\label{subsec:core_values}

We invited participants to share their understanding of privacy in the context of their design work. Our participants provided their perspectives on privacy, from motivation to practice, demonstrating how privacy is valued in relation to users, design goals, and organizational culture. We categorized these values as follows.

\subsubsection{Privacy as Data Protection Against Misuse}
For most participants, privacy was primarily framed as a technical matter to protect user data from unauthorized access and misuse. This perspective emphasizes safeguarding data through measures such as encryption, access controls, and legal compliance. As P7 noted, \textit{``it's about preventing or not allowing, essentially, like bad practices of aggregating data without user consent and making sure that if there is any input and information [about the user], you're doing the best... to have your designs prevent any form of misuse of the data.''} P9 described a similar approach using an example of designing an online resume platform, \textit{``Like PII, personally identifiable information, we have to make sure that this is stored securely and the channels through which it is sent this encrypted, and not shared across companies---like if the candidate is applying to Accenture, then the data is limited to Accenture and isn't shared with anyone else.''}

\subsubsection{Privacy as Part of User-Centered Design (UCD) Mindset}
Beyond technical means, another recurring theme among participants was the importance of placing users at the heart of design decisions---privacy is considered a part of this user-centered mindset and helps establish user trust. For example, as P12 described \textit{``We call it human-centered design, where the user is the main player here. Designs are humane and accessible and could be used by anyone and the user's privacy is maintained.''} This mindset is echoed by P5: \textit{``Design is about people. That's the motivator.''} Some designers illustrated how requesting unnecessary permissions could erode user trust, as P7 discussed, \textit{``It would put users at a kind of almost paranoid feeling to know that they [company] have access to this [camera functionality] when it's some kind of random app that wouldn't really need camera access.''} Participants frequently mentioned the need for empathy, \textit{``to think from their perspectives put ourselves in their shoes''} (P10), to consider the functional and emotional impact of their design choices. These perspectives indicate a collective desire of designers to champion user privacy and respect user boundaries. 

\subsubsection{Privacy as a Moral Obligation}
One theme in participants' perceptions of privacy was its framing as a moral obligation and an ethical responsibility to respect and protect users, even when for practices not explicitly defined within the scope of regulation. As P1 stated, \textit{``Ethical design is very big and privacy is part of it.''} This moral obligation might be defending users from potential harm from malicious online behaviour, such as doxxing and digital violence. It can also mean cultivating a healthy relationship with users, demonstrating the designers' care about users, their data, and their safety. P2 passionately described this aspiration to build such a relationship with users, \textit{``It wasn't a requirement, but we felt it was the right thing to do... We wanted to do better than the law because of this philosophy of caring about the people that we are creating stuff for.''}

\subsection{Factors Contributing to Designers' Values}
\label{subsec:factors}

Participants highlighted diverse factors that shape their understandings of and attitudes toward privacy, with particular attention to the social and collaborative contexts that support or hinder ethical development. We summarize these factors below.

\subsubsection{Motivated by Personal Experiences as a User}
Designers' personal experiences as users influenced their approach to privacy. Those who have encountered privacy violations or data misuse in their own lives mentioned those experiences and preferences as motivators for them to prioritize privacy in their work. 
P1 shared a vivid experience of trying to buy a coffee and was asked to provide multiple personal information, including her phone number. In this case, opting out or providing fake information was not an option because a phone number was required to receive the verification code. Once giving in, \textit{`` the same night I received so many SMS.''} (P1). A similar experience has been described by other participants. For example, P2 also described receiving unwelcome marketing on personal occasions: \textit{``I feel quite annoyed when I get random emails from companies on my birthday that they are trying to sell me something as they're pretending they wish me a happy birthday.''} P4 reflected on being part of a user community and how this stance impacts them as a designer, \textit{``We kind of know as a user---I don't want my information being taken away and sold to third-party companies... So a lot of it is navigating through our own moral ethics and putting ourselves as a user.''}

\subsubsection{Learning from Past Unethical Practices}
Encountering negative examples of ethical design practices in previous jobs has prompted some designers to become more vigilant about privacy concerns in their current work. Those examples motivated them to attend to the relationship between design and privacy, and start advocating for more responsible and transparent design practices. P1 recalled their experiences with dark patterns in their early work, \textit{``Back in the days, we didn't know what this thing actually is called dark pattern. So we actually did a lot of tricks to make people sign up for things.}'' After the awareness came into realization, P1 said: \textit{``And I started looking into it and realized this thing is called dark patterns... then I said to myself, you know, that is not right.''} In another example, P10 described how the emails to the users lack the unsubscribe option in their previous company on recruitment automation. Despite their team leader being aware of this issue, they took no action to correct it. As they gain more awareness of the regulative implication, they started to push back and challenge the top-down order: \textit{``later on only I realized that it's a big deal [of not having unsubscribe option] in some countries. People can actually sue the company if there is no option. ... We we have to tell them it's not possible. ''} (P10).

\subsubsection{Collaborative Influence in Privacy-Conscious Design}
\label{subsubsec:collaborative_positive}
At the same time, positive experiences also contributed to designers' privacy awareness. For example, P10 mentioned a mentor who encouraged integrating user privacy considerations into design: \textit{``My boss actually asked me to make sure that I consider user privacy as well in my designs. ... He asked me to look into user privacy, and that's when I started reading about it on the Internet.''} Since then, P10 started to understand and value the principles of minimizing data collection and transparency in their design. 
P2 also described how working with a business analyst and legal team broadened their perspective on privacy and decided to go \textit{``a little bit above and beyond the regulations.''}. Their extensive conversation on privacy and initiatives of performing privacy impact assessments motivated them to approach more privacy-conscious design \textit{``even a few years before the GDPR came into effect''} (P2).

\subsubsection{Exposure to Influential Figures and Resources}
\label{subsubsec:influential_figures}
Designers credited various external influences for shaping their understanding of privacy. Designers have been following advocates on privacy in design. P2 noted how Mike Monteiro, a design director, who fundamentally reshaped their mindset: \textit{``Mike Monteiro gave a talk about ethical design and how designers are messing up the world... and that for me was the big turning point.''}  
P4 also cited another figure, \textit{``Michael Bazzell---I think he's one of the best people working around privacy in the world. We use him as a foundation for our strategies, and he has a couple of books, very impactful...''} 
Designers also mentioned reading articles and doing self-directed research in the field in their learning process. For example, P10 shared, \textit{``When I started working with Figma and this user privacy thing, I went through some articles and I started following them. So at this point, I have some idea in my mind that when I'm working on a design, I have to make sure all these things are taken care of.''} P1 mentioned doing research, \textit{``Whenever I caught out company doing dodgy things, I have to do my own research... that gives me confidence [to recognize unethical practices]''} and then elaborated on their way of self research \textit{``I Google a lot and also I follow some people on LinkedIn... One of the organizations I follow is Fair Patterns. They talk about design patterns all the time, they also have their own term, their patterns and they do talk about privacy.''}

\subsubsection{Influence From Major Industrial Players}

Participants highlighted that their privacy attitudes and practices were often influenced by broader shifts within the industry, especially led by major corporations. For example, P12 mentioned: \textit{``If some big, big companies like Amazon and Meta take [privacy] as a practice... all the small companies will have to follow it.''} When major players adopt robust privacy frameworks as part of their core development process, smaller firms are more likely to imitate these practices, as P5 mentioned: \textit{``We based our privacy policy on theirs [big companies] and we tried to mitigate that as much as we can.''} As such, participants viewed the actions of major corporations as setting practical benchmarks that smaller companies, as well as individual designers, often referenced or adapted when developing their own privacy-related practices.

\subsection{Challenges to Address Privacy in Collaborative Environments}
\label{subsec:challenges}

We asked participants to discuss the challenges they faced when integrating privacy into their design process and to reflect on the obstacles during decision-making. Through our analysis, we spotted that many challenges lie in the organizational attitudes and the collaborative work environment toward privacy, where designers are forced to comply. Other factors, such as a lack of guidelines and resources, also hinder the integration of more privacy-enhanced design solutions.

\subsubsection{Conflict Between Business Goals and Privacy Considerations}

Many participants mentioned that they were forced to prioritize business objectives in their design, which may conflict with important privacy considerations. Business goals, particularly those tied to revenue generation, can take precedence, even when designers advocate for user-centric privacy practices. P5 explained this conflict, \textit{``You have designers advocating for the users, but the business has its specific goals that maybe they are financial goals and are very expensive to have the right solution, they sacrifice the best solution [for privacy].''} This happens more often with business models that heavily rely on data collection for revenue generation. P12 acknowledged this: \textit{``Sometimes certain business goals do not match with the privacy concern. Because if something is offered for free, there should be some way of revenue generation. So sometimes data is collected.''} P3 echoed this challenge in their work: \textit{``We have a very low retention rate, so we have to drive a lot of traffic through sales, through things like Google... Data privacy, to be honest, is part of it because it is bad. But at the end of the day, the focus is not on improving that... more so on how can we make the shop nicer and how can we offer better campaigns, you know, to drive more revenue.''}

The conflict between making revenue and protecting user privacy was typically raised from the adoption of third-party tracking and advertising. P12 illustrated their observation that the websites host Google personalized ads or any other ads through which a large amount of user data is collected, \textit{``sometimes when the company is starting and then it's a small-scale startup company, they need to create revenue, so they might need to collect some of the data.''} This conflict could also arise with the prioritization of certain features that are key for the business, where features can bring convenience to users (e.g., through personalization or automation) but inevitably compromise their privacy. P3 pointed this out: \textit{``Auto-suggestions which I think is a good thing because it makes it easier on the user when it comes to the checkout, because users don't have to manually put in information, but at the same time, it forces them to proceed because everything is there [pre-filled].''}

\subsubsection{Power Imbalance and Hierarchy Among Stakeholders}
Many participants expressed frustration over their limited influence on privacy decisions in the final product. While designers may advocate for user privacy and propose ethical solutions, the ultimate decision-making power often lies with other stakeholders, such as senior management, or even developers. These power dynamics created another barrier to implementing privacy into design. Just like P3 said, \textit{``You can drive awareness. You can make suggestions, but at the end of the day, if it's turned down [by others], it's turned down.''} P12 showed the fundamental power imbalance in decision making between the clients of the product and the freelance designers: \textit{``So sometimes data is collected... That depends on the decision maker, not the designer... Once it is handed over to the organization, it's under their control. I, as a freelance consultant, do not have much of a say in that.''} 

Many participants expressed concerns particularly when collaborating with developers to implement their design ideas for privacy protection. For example, P7 described their limited power, since the final product is implemented by the developers, they are not able to ensure their design is being developed, ``\textit{there's only so much that we could do as designers.''} Moreover, the designers are not able to control the entire data flow of the product, as P1 explained, \textit{``That is what I can do with UI, but what they [developers] do with the back end is something that I cannot control.''} Sometimes, the implementation constraints also limited the influence of the designers, as P4 suggested \textit{``After the handoff, we receive push-back from the developers like we can't do this feature because we have to collect whatever we have to call whatever API, or we need users to do single sign on, etc.''}

The level of experience of the designers can also affect the decision-making power of the designers. Particularly, entry-level designers have less impact, as their roles are typically execution-focused rather than strategic. P4 observed that many entry-level designers are not expected to discuss privacy concerns when \textit{``they get a brief from senior designers or managers and they just basically draw the shapes or make the flows.''} Designers \textit{``have a say''} on privacy policy when they are gaining more recognition in the organization and are involved in the business goals.  
On top of that, the size of companies and teams also matters, as P4 further explains, \textit{``You can always bring these issues up, but the likelihood of it having any impact when you're in a big company is like zero.''}

\subsubsection{Resource Constraints Within Organizations}
Many participants noted that privacy considerations are never prioritized due to heavy workload, tight deadlines, and competing responsibilities. For example, P7 emphasized, \textit{``There's not enough time for that [privacy considerations]. It's hard because we can only do so much within the time we have.''} P6, a solo designer, also described the challenge as relevant to user research: \textit{``for designers like me, don't have much time to really go to every user to ask their opinion and to do research.''} P1, explained the significant resource constraints in the context of being a design consultant in comparison with in-house designers, \textit{``In-house designers have more time to spend on things, but as a consultant, every minute is money, so your client obviously wants you to focus on something else instead of just the `terms and conditions' page.''} 
After all, resource limitations influence the priorities of the product. Privacy is often viewed as an optional feature rather than a core requirement, especially when companies are working under strict deadlines or financial constraints. P10 further illustrated the tension, \textit{``most of the time they [managers] worried about the timeline...the go-to-market thing. So if we keep telling that we need some more time to consider all this privacy-related stuff, they might reject that.''} P7 also highlighted that privacy consideration is regarded as \textit{``resource-heavy''}. Therefore, when the decision-making team are looking for corners to cut, \textit{``privacy might be the first thing that goes.''} As a designer working in a startup-sized company, P6 suggested that the tension is intensified: \textit{``We don't have a department or any person being tasked to focus on things like data privacy issues... It's just like we need to sacrifice sometimes some user experience just for faster development.''}

\subsubsection{Organizational Attention Limited to Legal Compliance}
\label{subsubsec:organization_challenges}
Participants mentioned that, rather than treating privacy as an ethical and UCD concern, companies often saw it through the narrow lens of legal compliance. As a result, they often only did the bare minimum to make their products comply with regulations. For example, P3 recounted an internal debate over implementing a cookie consent banner, \textit{``It has been even escalated to the high level like to the management... They said only do the minimal what is necessary.''} As such, privacy awareness tends to emerge not from an internalized value system, but from the fear of regulatory consequences. Participants reflected on past experiences where privacy considerations were largely ignored until potential legal implications came to light. For instance, when P10 described their early design of not having an unsubscribe option, P10 related, ``\textit{But later on only I realized that it's a big deal in some countries. ... People can actually sue the company if there is no option.}'' This risk-oriented mindset limited the proactive incorporation of privacy principles and can delay the recognition of critical issues until after harm or legal liability has already occurred. As P10 said: ``\textit{Serious things happen when maybe they'll do once and issues happen, so only then they [companies] will understand... Otherwise, they think that is just data.}''

\subsubsection{Lack of Clear and Actionable Guidelines}
\label{subsubsec:guidance_challenge}
Even with the need to comply with existing regulations, such as the GDPR, participants expressed concerns that there lacks clear and actionable guidelines that designers can easily interpret and implement. While these regulations set broad principles for data protection, participants found that they provide little concrete direction on how to integrate privacy considerations into the design process. For example, P3 noted the pain point of complying with GDPR: \textit{``You can read through GDPR, but it's more so written about that users should have access to their information and it should be mentioned how the data is stored. But to be honest, I have not come across good practice examples.''} In comparison, P3 found a much better support for adhering to other design principles, such as web accessibility, because they can easily find both good and bad design examples. 
The ambiguity in the regulations makes it difficult for designers to ensure compliance without additional legal consultation or internal policy development. As a result, many participants expressed that they were relying on their own knowledge or even gut feelings during their design process, as P12 noted,  \textit{``As a designer, I did not follow anything. It's whatever my understanding of whatever is sensitive information, I just followed that... that's whatever is in my knowledge.''}

\subsection{Adaptive Methods to Meet Challenges in Collaboration}
\label{subsec:adative_methods}
Our research identified several adaptive mechanisms designers employ in collaborative contexts with other stakeholders to protect user privacy while meeting business requirements. We summarize these approaches below.

\subsubsection{Proactively Integrating Privacy Early in the Process}
Participants emphasized the importance of embedding privacy considerations throughout the design process, including engaging users and making decisions about data collection from early on, and critically assessing what information is truly necessary to balance functionality with user trust. P7 described this as \textit{``from researching to actual designing and ideation''} and \textit{``in every stage of the [design] process''} rather than as an afterthought. 
P5 also suggested that the consideration of how to protect the users should start \textit{``as soon as you have a clear understanding about the problem that you are trying to solve.''}
To kickstart the privacy integration in the design process, participants mentioned the importance of establishing a clear data collection strategy. About this, P2 mentioned: \textit{``I first want to understand what kind of data we might be wanting to gather and then balancing that with the needs of people...''} 

To ensure that privacy concerns are consistently integrated in design and development, participants discussed the importance of documentation and tracing. On this, P12 mentioned, \textit{``In the design brief we set up a couple of design principles, regarding the ethical [including privacy] concerns.''} This approach ensures privacy is \textit{``kept in mind''} throughout development, as they further explained, \textit{``When the product is [being] developed, since this was a part of the design brief, it [privacy] will be taken into consideration.''}

\subsubsection{Challenging Status Quo and Raising Awareness}
Several participants actively sought to bring privacy awareness into their teams by directly challenging the existing practice or process, even when it was not formally part of their role. Echoing Section~\ref{subsec:core_values}, participants consider defiance in this case as their moral obligations. P1, as a design consultant, recounted challenging the existing data collection practice of their client company: \textit{``Do we really have to ask users to sign in the form before they even start a trial... The competitors are doing it and they just follow, they just don't think. And it is my job to tell them you don't have to ask users for their birthdays. If we don't need to ask, don't ask.''} P3 mentioned that even given the imbalance between business objectives and other considerations, they would still like to drive more awareness within their organization and have an impact on building trust with the users.

\subsubsection{Establishing Empathy and Mutual Understanding With Stakeholders}
Participants described using empathy and personal experiences to negotiate for privacy-friendly solutions with stakeholders without resorting to conflict. For example, P1, a design consultant, shared that they would recite their own experience when communicating with the clients and advocate privacy: \textit{``I just tell them, hey, the other day I went to a cafe. Someone asked me for my phone number. I just got really angry and yeah, I don't want this experience. I don't want your user to have this experience and yet usually people understand... Of course, I used a friendly tone to explain to them [my clients].''} 
P2 also emphasized negotiation through mutual understanding and alignment and described their most effective strategy as \textit{``employing a healthy dose of empathy.''} By learning the perspective of other stakeholders and understanding their strategies, P2 hoped to potentially offer \textit{``different tactics that still accomplish the strategies that they are trying to accomplish.''} P3 mentioned citing other sources (in their case, Baymard.com, a website providing evidence-based best practices for improving e-commerce UX design) for negotiation with others: \textit{``it shows best practices and worst practices, and explains why. That helps us when we go into discussion rounds with stakeholders. ... then we can show them precise examples, so it's easier to convince them.''}

\section{Discussion}

\begin{figure}
    \centering
    \includegraphics[width=0.99\linewidth]{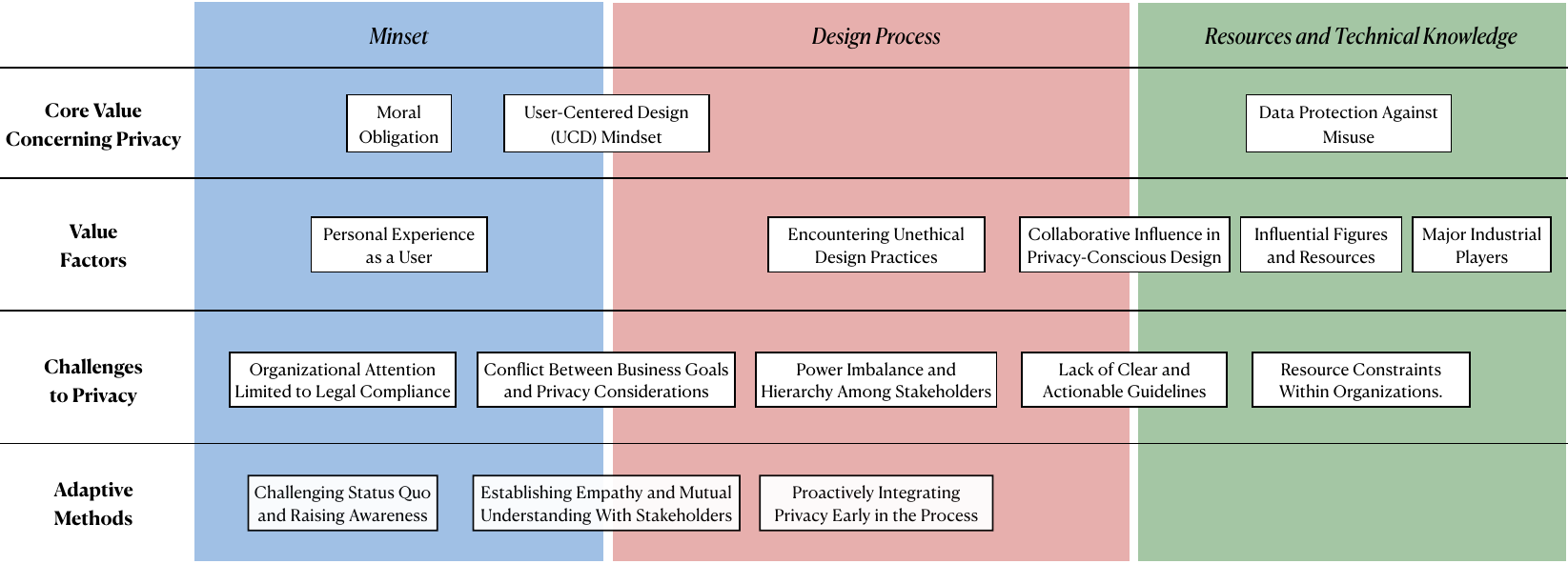}
    \caption{Our findings on the core values of UI/UX designers regarding privacy, the factors that shape their values, the challenges they face when integrating privacy considerations, and how they adapt to overcome these challenges. We further categorize those themes broadly into three aspects: \textit{mindset}, \textit{design process}, and \textit{resources and technical knowledge}, based on the primary emphasis of each theme, with some themes having overlapping emphases. }
    \label{fig:findings}
\end{figure}

Our study revealed privacy-advocating designers' core values concerning privacy, the social and collaborative factors shaping their perspectives, the challenges they face in integrating privacy in their professional work, and the adaptive methods they employ to meet some of those collaboration challenges. These values, factors, and  practical concerns are rooted in three contextual aspects, i.e., \textit{mindset}, \textit{design process}, and \textit{resources and technical knowledge} (summarized in Figure~\ref{fig:findings}). Our findings demonstrate that our participants hold strong ethical values and obligations to their users, but their ability to implement these values is constrained by organizational and collaborative factors rather than by their personal motivations or technical knowledge alone. Below, we contextualize our findings in broader research areas in CSCW and HCI and discuss their implications for the research community, design professionals, and privacy advocates.

\subsection{Unleash Designers' Privacy Advocacy Through UCD Principles}
Our study highlighted that, as users' strongest allies, UI/UX designers are ideally positioned to advocate for users' privacy. Many participants have already developed a deep empathy for users through their past experiences and the adoption of UCD methods. Therefore, considering the risks of privacy-invasive practices for users is a natural extension of their established design practices. Their attentiveness to poor design examples (such as the use of dark patterns), encountered either through their professional or personal lives, also provides them with a strong motivation for steering the design to more privacy-focused options. 

The question then arises: how can we foster awareness of privacy among designers in the first place? Many previous works consider privacy as a closely related concept to security~\cite{FactorsAffectingPrivacy, Evaluating_Privacy_Perceptions} and, therefore, are primarily concerned with technical means (e.g., encryption and access control). In such a framing, privacy is more relevant to the backend and the responsibility of developers. In contrast, our work implies that the connection between privacy and responsible design practice should be strengthened by privacy advocates. Both our findings and the previous work (such as the work by \citet{Chivukula1}) have demonstrated the potential link between UCD methods and privacy awareness, as well as more broadly, ethics awareness. It is very likely that the underlying principles for UCD help foster a user-centered mindset that is transferable to many complex scenarios that require the designers to have a moral and professional obligation to protect the users and respect their values and needs. Therefore, future research should further investigate the relationship between privacy awareness and UCD and explore how to adapt UCD methods to incorporate privacy-focused analysis (such as through value-sensitive design~\cite{friedman1996value}). 

\subsection{Bridge the Gap Between Motivation and Practical Impacts}
Eventually, pushing the privacy-focused design, and adopting UCD more broadly, into production requires all parties from the product team to be on board. However, our findings depict a persistent tension between business objectives and privacy considerations (the challenge of \textit{Conflict Between Business Goals and Privacy Considerations}). This tension is especially pronounced in companies offering free products that rely on data collection for revenue. Despite the current dominance of framing user data as an economic foundation, organizations need to recognize that privacy and business success are not inherently opposed. For example, companies can differentiate themselves by adopting privacy as a competitive advantage rather than viewing it as a source of cost. The privacy-focused strategy can also cultivate privacy innovations that open novel opportunities for organizations and industries. In the work by \citet{BLEIER2020466}, both aspects were illustrated with concrete examples (such as DuckDuckGo and blockchain-based products). Opening up such opportunities requires organizations to shift from compliance-focused approaches to considering privacy as a core value that enhances user trust and business in the long term.

Beyond these recommendations, we argue that we should push the changes further. Our study results highlight that the existing structures in various organizations exclude designers from the critical decision-making process. Given the essential role designers can play toward more ethical products, it is imperative to include designers in voicing their opinions and concerns at various stages of the technology design. This point has been echoed in many previous works in CSCW and HCI, such as \cite{Chivukula2,10.1145/3310276,Jones14122019}. Indeed, this is challenging, especially when diverse parties are involved in the decision-making process; there is no one-size-fits-all solution. Our study has illustrated the tension during decision-making resulting from our participants being either an external consultant, a freelancer, or a junior or single designer in their product team. Such design context should be carefully evaluated to understand how to promote shared and localized decision-making strategies.

\subsection{Create Designer-Centric Tools and Collaborative Platforms for Privacy}
As mentioned above, designers need to be empowered to have a stronger voice during collaboration in privacy-related decision-making. Many of the challenges raised by our participants can be potentially addressed by developing and adopting designer-centric processes and tools. Our findings highlight three types of tools promising to support designers' work.

First, designers' values around privacy often stem from their past experiences and the resources available to them, but they generally lack knowledge and effective resources to overcome integration challenges, as revealed in Figure~\ref{fig:findings}. Therefore, \textit{learning tools} that present case studies, synthesize best practices, and offer concrete guidance can raise privacy awareness and equip designers with the knowledge needed to approach ethical design and influence other stakeholders.

Second, our participants emphasized the need to proactively integrate privacy into the design process and noted the lack of actionable guidance. So, \textit{design tools} that directly support tracing privacy-related documentation, analysis of data flow, detection of privacy dark patterns, and suggestion of effective design solutions would streamline designers' work. Educational and instructional strategies for other established design principles, such as accessibility (as hinted by P3 in Section~\ref{subsubsec:guidance_challenge}), can be assessed and adopted for privacy. In addition, such tools can provide features allowing designers to exchange practices, templates, or annotated flows with each other; those features afford peer-to-peer support and collective sense-making, tapping into the positive value factors such as  \textit{Collaborative Influence in Privacy-Conscious Design} in Section~\ref{subsubsec:collaborative_positive}. 

Third, many of the challenges stemmed from organizational misconceptions and the marginalization of privacy concerns, as Section~\ref{subsubsec:organization_challenges} reveals. As a result, our participants made the effort to challenge the status quo and build mutual understanding across different duties and roles in organizations. In this context, \textit{workspace collaboration tools} that support evidence-based reasoning and provide shared discussion spaces to influence other stakeholders would help designers advocate for privacy more effectively, making it a team responsibility rather than an individual burden.

Beyond supporting individual teams and organizations, platforms that connect privacy-advocating designers with their peers, developers, legal experts, and broader stakeholder groups can positively impact their efficacy in designing privacy-enhanced solutions. Our results demonstrated the power of influential figures in encouraging designers and even organizations to follow the best practices (value factor of \textit{Influential Figures and Resources} in Section~\ref{subsubsec:influential_figures}). The communities of privacy advocates provide a space to share such knowledge, which would benefit the awareness and practical considerations for privacy. At the same time, our results indicate that the regulations concerning privacy are frequently vague and challenging to follow. Designers are typically not legal experts, and the legal texts seldom provide clear, actionable advice for integrating privacy into design workflows. Similar to GDPR and CCPA, local privacy laws are often limited to high-level principles and do not address the design concerns directly, which places an additional cognitive burden on practitioners~\cite{Horstmann_Domiks_Gutfleisch_Tran_Acar_Moonsamy_Naiakshina_2024}.  As evidenced by P2's experience (Section~\ref{subsubsec:collaborative_positive}), communities that enable the discussion on actionable guidelines with legal experts can help designers and other stakeholders establish common ground on privacy concerns and navigate the gray area between legal compliance and ethical responsibility. 

Together, these tools and platforms emphasize that promoting privacy is not only a matter of empowering individual designers but also about fostering collective efforts within and beyond existing organizational environments to make a systematic change toward privacy-sensitive design.

\subsection{Limitations}
Our study investigated UI/UX designers' perspectives on privacy through semi-structured interviews with 12 privacy-advocating design professionals. While this study provides valuable insights into how UI/UX designers perceive and implement privacy in their design practice, several limitations must be acknowledged. 

First, our participants are primarily composed of designers who have already demonstrated some level of awareness or interest in privacy issues. 
This decision was deliberate as we aimed to investigate the designers' understanding of privacy, what shaped their understanding, and how they negotiate for users' privacy in their work. Without a certain level of awareness of privacy, the participants would not be able to provide relevant information on those aspects. To make sure our participants fall into the preferred category, we started with the targeted recruitment of article authors on privacy and UI/UX design (in the first round). Other alternative methods~\cite{Robillard_trade_off_24} include recruiting through posting ads on public platforms and forums. After our initial investigation, such a method would take considerable effort to filter the candidate participants; it was particularly difficult to identify the designers with sufficient privacy awareness.  

Secondly, the sample size in our study was relatively small. The recruitment and analysis of our study were performed simultaneously---we started our analysis as soon as we completed each interview. While we continued analyzing the entire data until all interviews were completed, following the reflexive data analysis method, we observed that our understanding of the data and the derived themes stabilized after nine interviews. Therefore, we judge that our results are robust to meet our research objectives~\cite{Staller_sampling_21}. 

Finally, our findings were based on the reflections the participants verbalized during the interviews. While this is effective in deriving discursive knowledge and values of the participants, there might be a gap in the actual practice they deploy when encountering privacy concerns in the field. Moreover, we miss the perspective of other parties (such as developers) during their collaborative process. Future work can deploy research-through design and co-design methods to observe these factors in context.

\section{Conclusion}

Our study explored UI/UX designers' perceptions, values, and practices regarding privacy in collaborative settings through in-depth interviews with 12 privacy-advocating professional designers across diverse industries and regions. The findings highlight that these designers generally hold strong personal values about privacy, rooted in empathy and moral responsibility, and user-centered principles. Their ability to translate these values into practice, however, is shaped mostly by organizational structures and collaborative dynamics: organizational pressure to prioritize business goals, limited decision-making power due to stakeholder hierarchies, resource constraints, and unclear regulatory guidance. Despite these barriers, our participants adopt various strategies to negotiate privacy-conscious solutions, such as proactively integrating privacy early in the design process, challenging existing practices to raise awareness, and leveraging empathy and evidence-based discussions.

To bridge the gap between designers' motivations and practical outcomes, our work suggests the importance of empowering designers with clear guidelines, designer-centric privacy tools, and inclusion in strategic decision-making processes. Additionally, cultivating stronger interdisciplinary collaboration and building community among designers, developers, and business stakeholders can enhance collective privacy advocacy and practical implementation. Future research should focus on developing concrete, actionable frameworks and tools tailored specifically for UI/UX designers to streamline privacy integration and facilitate collaboration. Addressing these organizational and practical challenges can potentially help design teams more effectively champion user privacy, fostering trust and raising ethical standards in digital technology development.

\begin{acks}
    We thank our participants for their time and valuable insights. We also thank the anonymous reviewers for helping us improve the paper. This work is supported by Fonds de recherche du Québec (2022-PR-300101), the Canada Research Chairs program (CRC-2021-00076), and the Natural Sciences and Engineering Research Council of Canada (NSERC).
\end{acks}

\bibliographystyle{ACM-Reference-Format}
\bibliography{reference}
\end{document}